# Scheduling UET-UCT DAGs of Depth Two on Two Processors

Ruzayn Quaddoura
Computer Science Department
Zarqa University
Zarqa, Jordan
Ruzayn@zu.edu.jo

Gassan Samara
Computer Science Department
Zarqa University
Zarqa, Jordan
gsamara@zu.edu.jo

**Abstract—** *Given n unit execution time (UET) tasks whose precedence constraints form a directed acyclic graph (DAG), the arcs are associated with unit communication time (UCT) delays. The problem is to schedule the tasks on two processors in order to minimize the makespan. Several polynomial algorithms in the literature are proposed for special classes of digraphs, but the complexity of solving this problem in general case stills a challenging open question. We propose in this paper a linear time algorithm to compute an optimal schedule for the class of DAGs of depth two.*

## 1. Introduction

The problem of scheduling a set of tasks on a set of identical processors under a precedence relation has been studied for a long time. A general description of the problem is the following. There are $n$ tasks that have to be executed by $m$ identical processors subject to precedence constraints and (may be without) communication delays. The objective is to schedule all the tasks on the processors such that the makespan is minimum. Generally, this problem can be represented by a DAG $G = (V, E)$ called a task graph. The set of vertices $V$ corresponds to the set of tasks and the set of edges $E$ corresponds to the set of precedence constrains. A weight $p_i$ is associated with every vertex $i$ represents the execution time of the task $i$, and a weight $c_{ij}$ is associated with every edge $(i, j)$ represents the communication time between the tasks $i$ and $j$. If $(i,j) \in E$ and the task $i$ starts its execution at time $t$ on a processor $P$, then either $j$ starts its execution on $P$ at time greater than or equal to $t + p_j$, or $j$ starts its execution on some other processor at time greater than or equal to $t + p_j + c_{ij}$.

According to the three field notation scheme introduced in [11] and extended in [1] for scheduling problems with communication delays, this problem is denoted as $P_m|prec, p_i, c_{ij}|C_{max}$.

A large amount of work in the literature studies this problem with a restriction on its structure: the time of execution of every task is one unit execution time (UET), the number of processors $m$ is fixed, the communication delays are neglected, constant or one unit (UCT), or special classes of task graph are considered.
We find In this context, the problem $P_2|prec, p_i = 1|C_{max}$, is polynomial [2, 6], i.e. when the communication delays are not taken into account. On the contrary the problem $P_3|prec, p_i = 1|C_{max}$ remains an open question [8].

The problem of two processors scheduling with communication delays is extensively studied [7, 10]. In particular, it is proven in [3] that the problem $P_2|prec = binary\ tree, p_i = 1, c_{ij} = c|C_{max}$ is NP-hard where $c$ is a large integer, whereas this problem is polynomial when the task graph is a complete binary tree.

A challenging open problem is the two processors scheduling with UET-UCT, i.e. the problem $P_2|prec, p_i = 1, c_{ij} = 1|C_{max}$ for which the complexity is unknown. However, several polynomial algorithms have been shown for special classes of task graphs, especially for trees [9, 13], interval orders [1] and a sub-class of series parallel digraphs [5]. In this paper we propose linear time algorithm to compute an optimal scheduling for the class of DAGs of depth two, that is the digraphs for which the set of vertices can be partitioned into three levels $A, B$ and $C$ where $A$ is the set of sources (vertices without predecessors), $C$ is the set of sinks (vertices without predecessors), and $B$ is the set of vertices located between $A$ and $C$, that is, for any vertex $x \in B$, $x$ has a predecessor in $A$ and has or hasn't a successor in $C$, in addition there is no edge between any vertex of $A$ and any vertex of $C$. Such DAG will be denoted by $G = (A \cup B \cup C, E)$. This algorithm is an extension of the $O(n)$ time algorithm for the class of bipartite digraphs of depth one presented in [12].

## 2. Preliminaries

A schedule UET-UCT on two processors for a general DAG $G = (V, E)$ is defined by a function $\sigma: V \rightarrow \mathbb{N}^+ \times \{P_1, P_2\}$, $\sigma(v) = (t_v, P_i), i = 1,2$ where $t_v$ is the time for which the vertex $v$ is executed and $P_i$ the processor on which the vertex $v$ is scheduled. A schedule $\sigma$ is feasible if :

a) $\forall u, v \in V$, if $u \neq v$ then $\sigma(u) \neq \sigma(v)$
b) If $(u, v) \in E$ then $t_u + 1 \leq t_v$ if $u$ and $v$ are scheduled on the same processor, and $t_u + 2 \leq t_v$ if $u$ and $v$ are scheduled on distinct processors.

A time $t$ of a schedule $\sigma$ is said to be idle if one of the processors is idle during this time. The makespan $C_{max}$ or the length of a schedule $\sigma$ is the last non-idle time of $\sigma$, that is :

$$C_{max} = \max\{t : \exists\ v \in V\ \sigma(v) = (t, P_i), i = 1\ or\ 2\}$$

A schedule $\sigma$ is optimal if $C_{max}$ is the minimum among all feasible schedules.

Let $G = (A \cup B \cup C, E)$ be a DAG of depth two. There exists always a feasible schedule for which the set of sources

$A$ are executed first and then vertices of B and finally the set of sinks C. Our algorithm for solving the problem under consideration produces an optimal schedule satisfies this condition and that we called a natural schedule defined as following:

**Definition 1** Let $G = (A \cup B \cup C, E)$ be a DAG of depth two. A natural schedule of $G$ is obtained by scheduling in order first the sources A then the vertices B and finally the sinks C starting from the processor $P_1$ and alternating between $P_1$ and $P_2$ such that the resulting schedule is optimal.

The definition of a natural schedule $\sigma$ of a DAG $G = (A \cup B \cup C, E)$ of depth two implies the following property:
$$\lceil |A \cup B \cup C|/2 \rceil \leq C_{max} \leq \lceil |A \cup B \cup C|/2 \rceil + 2$$

The justification of these inequalities is the following: If the processors $P_1$ and $P_2$ must be idle between the vertices of $A$ and $B$, also must be idle between $B$ and $C$ then $C_{max} = \lceil |A \cup B \cup C|/2 \rceil + 2$, otherwise, $C_{max} = \lceil |A \cup B \cup C|/2 \rceil + 1$ or $C_{max} \leq \lceil |A \cup B \cup C|/2 \rceil$. Our scheduling algorithm produces a natural scheduling satisfies this property.
A set of vertices $X$ is called stable set if there is no edge between any two vertices of $X$. By definition, in a DAG of depth two, $G = (A \cup B \cup C, E)$, the sets $A, B$ and $C$ are stable sets.
Let $X$ be a stable set of vertices. Suppose the processor $P_1$ or $P_2$ is ready to execute the vertices of $X$ starting at some time t. The set X is called of type even (resp. odd) if one of the following is hold:
1) Starting of time $t$, $P_1$ and $P_2$ are both ready to execute the vertices of $X$ and $|X|$ is even (resp. odd)
2) Starting of time $t$, $P_1$ only is ready to execute the vertices of $X$ and $|X|$ is odd (resp. even)

Remark that, if a set $X$ is of type even then the last two scheduled vertices of $X$ are executed in parallel on $P_1$ and $P_2$ at the same time. These two vertices are called a vertical hinge of $X$. If $X$ is of type odd then the last two scheduled vertices of $X$ are executed in series on $P_1$. These two vertices are called a horizontal hinge of $X$.
Let $x$ be a vertex of a DAG $G = (A \cup B \cup C, E)$ of depth two. The positive neighbors set of $x$ is the set $N^+(x) = \{y : (x,y) \in E\}$. The positive degree of $x$ denoted by $d^+(x) = |N^+(x)|$. The negative neighbors set of $x$ is the set $N^-(x) = \{y : (y,x) \in E\}$. The negative degree of $x$ denoted by $d^-(x) = |N^-(x)|$. Remark that, for any $x \in A, N^-(x) = \emptyset$, and may be $N^+(x)$ empty or not. For any $x \in C, N^+(x) = \emptyset$ but $N^-(x)$ can't be empty. If $x \in B$, $N^-(x) \neq \emptyset$ and may be $N^+(x)$ empty or not. The non neighbors set of $x$ is the set $\overline{N}(x) = \{y : (x,y) \text{ and } (y,x) \notin E\}$. Let $X$ be a set of vertices, the sub graph induced by $X$ is denoted by $G[X]$.

## 3. Scheduling algorithm

The general idea of solving the problem $P_2|prec = (A \cup B \cup C, E), p_j = 1, c_{ij} = 1|C_{max}$ is to determine the favorable last two vertices of $A$ say $a_1, a_2$ that must be scheduled before scheduling the vertices of $B$ and the favorable first two vertices of $B$ say $b_1, b_2$ that must be scheduled directly after $a_1, a_2$. Then, determine the favorable last two vertices of $B$ say $b_3, b_4$ that must be scheduled before scheduling the vertices of $C$ and the favorable first two vertices of $C$ say $c_1, c_2$ that must be scheduled directly after $b_3, b_4$. The essential mission of our algorithm is the decision whether the vertex $b_1$ or $b_2$ and $b_3$ or $b_4$ exist or not and the decision whether the vertex $c_1$ or $c_2$ exist or not.
For technical reasons of the work of the algorithm, it produces an optimal natural scheduling in reverse, that is, it starts to schedule first the set of sinks $C$ then B and finally the set of sources A. Obviously, if we read this producing optimal scheduling in reverse, we obtain the demanded optimal scheduling.
For a clear reading, we will summarize the grand lines of our algorithm as following.

**Grand lines of Scheduling Algorithm**
1) If $|C|$ is even, the algorithm search of the two vertices of $C$ say $(c_1, c_2)$ that must be scheduled at time $\frac{|C|}{2}$ on $P_1$ and $P_2$, and the two vertices of $B$ say $(b_1, b_2)$ that must be scheduled at time $\frac{|C|}{2} + 1$ on $P_1$ and $P_2$. It is possible that $b_2$ can't be exist or $b_1$ and $b_2$ can't be exist. In this case, the algorithm consider that there is an imaginary vertex denoted by $i$ in place of $b_2$ or in place of $b_1$ and $b_2$. Since the vertices $c_1$ and $c_2$ will be scheduled in parallel on $P_1$ and $P_2$, as well $b_1$ and $b_2$, the ordered pair $(c_1, c_2)$ is called a Vertical Right Hinge of $C$ and the ordered pair $(b_1, b_2)$ is called a Vertical Left Hinge of $B$.
2) If $|C| > 1$ is odd, the algorithm search of the two vertices of $C$ say $(c_1, c_2)$ that must be scheduled at times $\lceil \frac{|C|}{2} \rceil - 1$ and $\lceil \frac{|C|}{2} \rceil$ on $P_1$, and the two vertices of $B$ say $(b_1, b_2)$ that must be scheduled at times $\lceil \frac{|C|}{2} \rceil$ and $\lceil \frac{|C|}{2} \rceil + 1$ on $P_2$. In this case, it is possible to be $b_1$ or $b_2$ an imaginary vertex. Since the vertices $c_1$ and $c_2$ will be scheduled in series on $P_1$, as well $b_1$ and $b_2$ will be scheduled in series on $P_2$, the ordered pair $(c_1, c_2)$ is called a Horizontal Right Hinge of $C$ and the ordered pair $(b_1, b_2)$ is called a Horizontal Left Hinge of $B$. If $|C| = 1$, the algorithm schedule $C = \{c_1\}$ on $P_1$ and search of a Horizontal Left Hinge $(b_1, b_2)$ of $B$ that must be scheduled at times 1 and 2 on $P_2$.
3) If $|B| > 3$ then $B = B - \{b_1, b_2\}$ else the algorithm takes into consideration all vertices of $B$.
4) If the hinge $(b_1, b_2)$ is vertical then according to type $|B|$ either even or odd, the algorithm repeats step 1 or 2 for determining the right hinge $(b_3, b_4)$ of $B$ and the left hinge $(a_1, a_2)$ of $A$. If $|B|$ is even then $(b_3, b_4)$ and $(a_1, a_2)$ are vertical, in this case, it is possible that $a_2 = i$ or $a_1 = a_2 = i$. If $|B|$ is odd then $(b_3, b_4)$ and $(a_1, a_2)$ are horizontal, in this case, it is possible that $a_1 = i$ or $a_2 = i$. The scheduling time for $(b_3, b_4)$ and $(a_1, a_2)$ will be determined through the code of the algorithm.

5) If the hinge $(b_1, b_2)$ is horizontal then, $B = B - \{b_1, b_2\}$ can't be empty. This is obvious if $b_1 = i$ or $b_2 = i$. Also this is obvious if $b_1 \neq i$ and $b_2 \neq i$, otherwise, since $c_1 b_1, c_1 b_2 \notin E$ where $c_1$ is the first component of the horizontal right hinge of $C$, then $c_1$ is an isolated sink, so $c_1$ does not belong to $C$, contradiction. Suppose $|B| = |B - \{b_1, b_2\}| > 1$, if $|B|$ is even, then the algorithm consider that $B$ is of type odd and repeats step 2 to find the horizontal left hinge $(b_3, b_4)$ of $B$ and the horizontal right hinge $(a_1, a_2)$ of $A$. If $|B|$ is odd, then the algorithm consider that $B$ is of type even and repeats step 1 to find the vertical left hinge $(b_3, b_4)$ of $B$ and the vertical right hinge $(a_1, a_2)$ of $A$. Suppose $|B| = |B - \{b_1, b_2\}| = 1$, if $b_2 \neq i$, then the algorithm consider that $B$ is of type even and repeats step 1 to find the vertical left hinge $(b_3, b_4)$ of $B$ and the vertical right hinge $(a_1, a_2)$ of $A$. If $b_2 = i$ then the algorithm consider that $B$ is of type odd and repeats step 2 to find the horizontal left hinge $(b_3, b_4)$ of $B$ and the horizontal right hinge $(a_1, a_2)$ of $A$. The scheduling time for $(b_3, b_4)$ and $(a_1, a_2)$ will be determined through the code of the algorithm.

6) The scheduling times for the vertices $C - \{c_1, c_2\}$ will be smaller than or equal to the scheduling times of $\{c_1, c_2\}$, the scheduling times for the vertices $B - \{b_1, b_2, b_3, b_4\}$ will be between the scheduling times of $\{b_1, b_2\}$ and $\{b_3, b_4\}$, and scheduling times for the vertices $A - \{a_1, a_2\}$ will be greater than or equal to the scheduling times of $\{a_1, a_2\}$.

Figure 1 illustrates a DAG $G = (A \cup B \cup C, E)$ of depth two and its reverse natural scheduling. For this example, V.R.hinge( $C$ ) = $(c_3, c_4)$, V.L.hinge( $B$ ) = $(b_4, i)$, H.R.hinge($B$) = $(b_2, b_1)$, H.L.hinge($A$ ) = $(a_3, a_2)$.

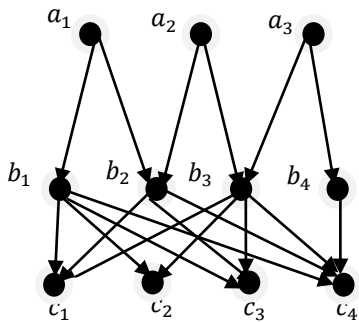

| $P_1$ | $c_1$ | $c_4$ | $b_4$ | $b_2$ | $b_1$ | $a_1$ |
| $P_2$ | $c_2$ | $c_3$ |       | $b_3$ | $a_3$ | $a_2$ |
|       | 1     | 2     | 3     | 4     | 5     | 6     |

Figure 1 A DAG of depth two and its reverse natural optimal scheduling

Let's present now the mathematical details of steps one to six. The discussion above show that the building of optimal scheduling for a DAG $G = (A \cup B \cup C, E)$ of depth two, starts by searching first the right hinge of $C$ and the left hinge of $B$ in $G[B \cup C]$, then we repeat this work for $G[A \cup B]$ or for $G[A \cup (B - (b_1, b_2))]$. So we will starts to build useful tools concerning this search for a bipartite graph of depth one $G = (B \cup W, E)$, i.e. $B$ is the set of sources and $W$ is the set of sinks. We need the following two Lemmas which are proved in [12].

**Lemma 2** Let $G = (B \cup W, E)$ be a bipartite graph of depth one. Assume that $|W|$ is even.
1. The two processors $P_1$ and $P_2$ are idle at time $\frac{|W|}{2} + 1$ if and only if $G$ is a bipartite complete.
2. The processor $P_2$ only is idle at time $\frac{|W|}{2} + 1$ if and only if one of the following holds:
   a. There is exactly one vertex $b \in B$ such that $d^+(b) < |W|$
   b. There is exactly one vertex $w \in W$ such that $d^-(w) < |B|$

**Lemma 3** Let $G = (B \cup W, E)$ be a bipartite graph of depth one. Assume that $|W|$ is odd.
1. The processor $P_2$ is idle at times $\lceil |W|/2 \rceil$ and $\lceil |W|/2 \rceil + 1$ if and only if $G$ is a bipartite complete.
2. The processor $P_2$ is idle at time $\lceil |W|/2 \rceil$ and not idle at time $\lceil |W|/2 \rceil + 1$ if and only if for every $b \in B$, $d^+(b) \geq |W| - 1$.
3. The processor $P_2$ is idle at time $\lceil |W|/2 \rceil + 1$ and not idle at time $\lceil |W|/2 \rceil$ if and only if
   a. There is $b \in B$ such that $d^+(b) \leq |W| - 2$
   b. If $b \in B$ such that $d^+(b) \leq |W| - 2$ then for every $w \in \bar{N}_W(b)$ $d^-(w) = |B| - 1$ where $\bar{N}_W(b)$ is the set of non neighbors of $b$ in $W$.

We need to translate the condition 2.b in Lemma 2 as well as conditions 2 and 3 in Lemma 3 in terms of the vertices of $W$.

**Lemma 4** Let $G = (B \cup W, E)$ be a bipartite graph of depth one. The following statements are equivalent:
1. There is exactly one vertex $b \in B$ such that $d^+(b) < |W|$.
2. For every $w \in W, d^-(w) \geq |B| - 1$, and if $w_1, w_2 \in W$ such that $d^-(w_1) = d^-(w_2) = |B| - 1$ then $N^-(w_1) = N^-(w_2)$

**Proof** $1 \Rightarrow 2$: If there is $w \in W$ such that $d^-(w) \leq |B| - 2$, then there are $b_1, b_2 \in B$ such that $b_1 w, b_2 w \notin E$, thus, $d^+(b_1) < |W|$ and $d^+(b_2) < |W|$, contradiction.
$2 \Rightarrow 1$: Suppose that there are $b_1, b_2 \in B$ such that $d^+(b_1) < |W|$ and $d^+(b_2) < |W|$. If $N^+(b_1) \subseteq N^+(b_2)$ then, since $d^+(b_2) < |W|$, there is $w \notin N^+(b_2)$, so $w \notin N^+(b_1)$, thus

$d^-(w) \leq |B| - 2$, contradiction. Suppose there is $w_1 \in N^+(b_1) - N^+(b_2)$ and $w_2 \in N^+(b_2) - N^+(b_1)$ then according to supposition, $d^-(w_1) = d^-(w_2) = |B| - 1$. Thus, $N^-(w_1) = B - \{b_2\}$ and $N^-(w_2) = B - \{b_1\}$, but now, $N^-(w_1) \neq N^-(w_2)$, contradiction.

**Lemma 5** Let $G = (B \cup W, E)$ be a bipartite graph of depth one. The following statements are equivalent:
1. For every $b \in B$, $d^+(b) \geq |W| - 1$.
2. For every $w_1, w_2 \in W$, $\bar{N}_B(w_1) \cap \bar{N}_B(w_2) = \emptyset$

**Proof** 1 ⇒ 2: If there are $w_1, w_2 \in W$ such that $\bar{N}_B(w_1) \cap \bar{N}_B(w_2) \neq \emptyset$ then for any $\in \bar{N}_B^-(w_1) \cap \bar{N}_B^-(w_2)$, $d^+(b) \leq |W| - 2$, contradiction.
2 ⇒ 1: If there is $b \in B$ such that $d^+(b) \leq |W| - 2$ then, there are $w_1, w_2 \in W$ and $b \in \bar{N}_B(w_1) \cap \bar{N}_B(w_2)$, contradiction.

**Lemma 6** Let $G = (B \cup W, E)$ be a bipartite graph of depth one. The following statements are equivalent:
1. If $b \in B$ such that $d^+(b) \leq |W| - 2$ then for every $w \in \bar{N}_W(b)$, $d^-(w) = |B| - 1$.
2. For every $w_1, w_2 \in W$, $|\bar{N}_B(w_1) \cap \bar{N}_B(w_2)| \leq 1$ and if $|\bar{N}_B(w_1) \cap \bar{N}_B(w_2)| = 1$ then $d^-(w_1) = d^-(w_2) = |B| - 1$

**Proof** 1 ⇒ 2: By Lemma 5, if for every $b \in B$, $d^+(b) \geq |W| - 1$, then for every $w_1, w_2 \in W$, $|\bar{N}_B(w_1) \cap \bar{N}_B(w_2)| = 0$. Suppose there is $b \in B$ such that $d^+(b) \leq |W| - 2$. Let $w_1, w_2 \in W$. If $w_1, w_2 \in \bar{N}_W(b)$ then by supposition $d^-(w_1) = d^-(w_2) = |B| - 1$, thus, $N^-(w_1) = N^-(w_2) = B - \{b\}$, hence $|\bar{N}_B(w_1) \cap \bar{N}_B(w_2) = \{b\}$.
If $w_1 \in N^+(b)$ and $w_2 \notin N^+(b)$ i.e. $w_2 \in \bar{N}_W(b)$, then $d^-(w_2) = |B| - 1$, thus $N^-(w_2) = B - \{b\}$, so $\bar{N}_B(w_2) = \{b\}$. Now, $\bar{N}_B(w_1) \cap \bar{N}_B(w_2) = \emptyset$.
If $w_1, w_2 \in N^+(b)$ then, if $\bar{N}_B(w_1) \cap \bar{N}_B(w_2) = \emptyset$, there is nothing to prove. Suppose $\bar{N}_B(w_1) \cap \bar{N}_B(w_2) \neq \emptyset$, let $b' \in \bar{N}_B(w_1) \cap \bar{N}_B(w_2)$, then $d^+(b') \leq |W| - 2$, since $w_1, w_2 \in \bar{N}_W(b')$ then by supposition, $d^-(w_1) = d^-(w_2) = |B| - 1$, so $\bar{N}_B(w_1) \cap \bar{N}_B(w_2) = \{b'\}$.
2 ⇒ 1: Let $b \in B$ such that $d^+(b) \leq |W| - 2$, let $w \in \bar{N}_W(b)$. Since $d^+(b) \leq |W| - 2$, there is $w' \in \bar{N}_W(b)$ and $w' \neq w$. by supposition $|\bar{N}_B(w) \cap \bar{N}_B(w')| \leq 1$. Since $b \in \bar{N}_B(w) \cap \bar{N}_B(w')$ then $|\bar{N}_B(w) \cap \bar{N}_B(w')| = 1$, thus by supposition, $d^-(w) = d^-(w') = |B| - 1$.

We are ready now to translate Lemmas 2 and 3 in terms of vertical hinge, horizontal hinge and the vertices of $W$ in a bipartite DAG of depth one $G = (B \cup W, E)$. According to Lemmas 4, 5, and 6 we have the following Corollaries.

**Corollary 7** Let $G = (B \cup W, E)$ be a bipartite graph of depth one. Assume that $|W|$ is even.
1. The vertical left hinge of $B$ is $(i, i)$ if and only if $G$ is a bipartite complete.
2. The vertical left hinge of $B$ is $(b_0, i)$, where $b_0 \in B$, if and only if one of the following holds:
   a. For every $w \in W$, $d^-(w) \geq |B| - 1$, and if $w_1, w_2 \in W$ such that $d^-(w_1) = d^-(w_2) = |B| - 1$ then $N^-(w_1) = N^-(w_2)$.
   b. There is exactly one vertex $w \in W$ such that $d^-(w) < |B|$.

**Corollary 8** Let $G = (B \cup W, E)$ be a bipartite graph of depth one. Assume that $|W|$ is odd.
1. The horizontal left hinge of $B$ is $(i, i)$ if and only if $G$ is a bipartite complete.
2. The horizontal left hinge of $B$ is $(i, b_0)$, where $b_0 \in B$, if and only if for every $w_1, w_2 \in W$, $\bar{N}_B(w_1) \cap \bar{N}_B(w_2) = \emptyset$
3. The horizontal left hinge of $B$ is $(b_0, i)$, where $b_0 \in B$, if and only if for every $w_1, w_2 \in W$, $|\bar{N}_B(w_1) \cap \bar{N}_B(w_2)| \leq 1$ and if $|\bar{N}_B(w_1) \cap \bar{N}_B(w_2)| = 1$ then $d^-(w_1) = d^-(w_2) = |B| - 1$

If the hinge of $G = (B \cup W, E)$ does not contains the imaginary vertex $i$ then the length of the optimal scheduling $\sigma$ is $\lceil |B \cup W|/2 \rceil$.

**Corollary 9** Let $G = (B \cup W, E)$ be a bipartite graph of depth one and $|W|$ is even. $|\sigma| = \lceil |B \cup W|/2 \rceil$ if and only if there are $w_1, w_2 \in W$ such that $d^-(w_1) \leq |B| - 1$, $d^-(w_2) \leq |B| - 1$ and $N^-(w_1) \neq N^-(w_2)$.
**Proof** The condition of Corollary is the negation of condition 2.b in Lemma 2. Also it is the negation of condition 2.a, otherwise if there is exactly one vertex $b \in B$ such that $d^+(b) < |W|$ then, by Lemma 4, for every $w \in W$, $d^-(w) \geq |B| - 1$, so $d^-(w_1) = |B| - 1$, $d^-(w_2) = |B| - 1$, thus by Lemma 4, $N^-(w_1) \neq N^-(w_2)$, contradiction.

The negation of Lemma 6 implies the following Corollary.

**Corollary 10** Let $G = (B \cup W, E)$ be a bipartite graph of depth one and $|W|$ is odd. $|\sigma| = \lceil |B \cup W|/2 \rceil$ if and only if one of the following is hold
1. There are $w_1, w_2 \in W$ such that $|\bar{N}_B(w_1) \cap \bar{N}_B(w_2)| \geq 2$
2. There are $w_1, w_2 \in W$ such that $|\bar{N}_B(w_1) \cap \bar{N}_B(w_2)| = 1$ and ($d^-(w_1) \neq |B| - 1$ or $d^-(w_2) \neq |B| - 1$).

We will translate now the Corollaries 6, 7, 8, and 9 to pseudo codes Procedure Vertical Hinge($G = (B \cup W, E)$) return a vertical right hinge of $W$ (V.R.H($W$)) and a vertical left hinge of $B$ (V.L.H($B$)). Corollaries 7 and 9 prove Its correctness.

**Procedure Vertical Hinge($G = (B \cup W, E)$)**
$W_1 = \{w \in W : d^-(w) = |B|\}$
$W_2 = \{w \in W : d^-(w) < |B|\}$
If $W_1 = W_2$ then
  Let $w_1, w_2$ any two vertices of $W$, V.R.H ($W$) = $(w_1, w_2)$ and V.L.H ($B$) = $(i, i)$
Else if $W_2 = \{w_0\}$ then
  Let $b_0 \in \bar{N}_B(w_0)$, let $w \in W_1$, V.R.H ($W$) = $(w, w_0)$ and V.L.H ($B$) = $(b_0, i)$

Else if there is $b_0 \in B$ such that for every $w_1, w_2 \in W_2$, $N^-(w_1) = N^-(w_2) = B - \{b_0\}$ then
  Let $w_1 \in W$ and $w_2 \in W_2$, V.R.H $(W) = (w_1, w_2)$, V.L.H $(B) = (b_0, i)$
Else let $w_1, w_2 \in W_2$ and $b_1, b_2 \in B$ such that $(b_1, w_2), (b_2, w_1) \notin E$, V.R.H$(W) = (w_1, w_2)$, V.L.H $(B) = (b_1, b_2)$
Return V.R.H $(W)$ and V.L.H$(B)$

Procedure Horizontal Hinge( $G = (B \cup W, E)$ ) return a horizontal right hinge of $W$, H.R.H$(W)$ and horizontal left hinge of $B$, H.L.H$(B)$. Corollaries 8 and 10 prove Its correctness.

**Procedure Horizontal Hinge($G = (B \cup W, E)$)**
$W_1 = \{w \in W: d^-(w) = |B|\}$
$W_2 = \{w \in W: d^-(w) \leq |B| - 1\}$
$W_3 = \{w \in W: d^-(w) \leq |B| - 2\}$
If $|W| = 1$ then let $W = \{w_0\}$
  if $N^-(w_0) = B$ then H.R.H $(W) = w_0$ and H.L.H $(B) = (i, i)$ else let $b_0 \in B$ such that $b_0 w_0 \notin E$, H.R.H $(W) = w_0$ and H.L.H $(B) = (b_0, i)$
Else if $W_1 = W$ then
  Let $w_1, w_2$ any two vertices of $W$, H.R.H $(W) = (w_1, w_2)$ and H.L.H $(B) = (i, i)$
Else let $W_1 = \{w_1, \dots w_r\}, W_2 = \{w_{r+1}, \dots, w_s\}$
  For $i = 1$ to $r$
    Let $\bar{N}_B(w_i) = \{b\}$
    If $b$ is not marked then Mark $(b) = \{w_i\}$ else Mark$(b)$=Mark$(b) \cup \{w_i\}$
  For $i = r + 1$ to $s$
    Let $\bar{N}_B(w_i) = \{b_1, \dots, b_k\}$
      For $j = 1$ to $k$
        If $b_j$ is not marked then Mark$(b_j)= \{w_i\}$
        Else let $w \in$ Mark $(b_j)$ and $b \in \bar{N}_B(w_i)$, H.R.H$(W)=(w, w_i)$ and H.L.H$(B) = (b_j, b)$
  For $i = 1$ to $r$
    If $|\text{Mark}(\bar{N}_B(w_i))| > 1$ then
    let $w \in \bar{N}_B(w_i)$ and $b \in \bar{N}_B(w_i) \cap \bar{N}_B(w)$ H.R.H $(W) = (w_i, w)$ and H.L.H $(B) = (b, i)$
  Let $w_1, w_2 \in W$ and $b \in \bar{N}_B(w_2)$, H.R.H$(W) = (w_1, w_2)$ and H.L.H $(B) = (i, b)$
  Return H.R.H$(W)$ and H.L.H $(B)$.

We are ready now to present the main result represented by the following scheduling algorithm for a DAG $G = (A \cup B \cup C, E)$ of depth two.

**Scheduling algorithm**
If $|C|$ is even then
  Vertical Hinge($G[B \cup C]$)
  $(c_1, c_2) = $ V. R. H $(C)$
  $(b_1, b_2) = $ V. L. H $(B)$
Else Horizontal Hinge($G[B \cup C]$)
  $(c_1, c_2) = $ H. R. H $(C)$
  $(b_1, b_2) = $ H. L. H $(B)$
If $|B| > 3$ then $B = B - \{b_1, b_2\}$
If $(b_1, b_2) = $ V. L. H $(B)$ then
  If $|B|$ is even then
    Vertical Hinge($G[B \cup A]$)
    $(b_3, b_4) = $ V. R. H $(B)$
    $(a_1, a_2) = $ V. L. H $(A)$
  Else Horizontal Hinge($G[B \cup A]$)
    $(b_3, b_4) = $ H. R. H $(B)$
    $(a_1, a_2) = $ H. L. H $(A)$
Else //$(b_1, b_2) = $ H. L. H $(B)$
if $|B| > 1$ then
  If $|B|$ is even then
    Horizontal Hinge($G[B \cup A]$)
    $(b_3, b_4) = $ H. R. H $(B)$
    $(a_1, a_2) = $ H. L. H $(A)$
  Else Vertical Hinge($G[B \cup A]$)
    $(b_3, b_4) = $ V. R. H $(B)$
    $(a_1, a_2) = $ V. L. H $(A)$

Else let $B = \{b_3\}$
  If $b_2 \neq i$ then
    Vertical Hinge($G[\{b_2, b_3\} \cup A]$)
    $(b_3, b_4) = $ V. R. H $(\{b_2, b_3\})$
    $(a_1, a_2) = $ V. L. H $(A)$
  Else Horizontal Hinge($G[\{b_3\} \cup A]$)
    $b_3 = $ H. R. H $(\{b_3\})$
    $(a_1, a_2) = $ H. L. H $(A)$

Schedule $A, B, C$ alternately on $P_1$ and $P_2$ in the order $C - (c_1, c_2), (c_1, c_2), (b_1, b_2), B - (b_1, b_2), (b_3, b_4), (a_1, a_2), A - (a_1, a_2)$

**Complexity**
We can verify that the proposed scheduling algorithm run in $O(n + m)$ time where $n = |A \cup B \cup C|$ and $m = |E|$.

## 4. Conclusion
We have proposed an $O(n + m)$ time algorithm for the optimal scheduling UET-UCT DAGs of depth two on two processors. The complexity of this problem for general directed acyclic graphs is still an open question. We believe that our algorithm can be used to solve this problem in general as follow: Consider a topological sort of a directed acyclic graph G. The linear ordering defined by this topological sort decomposes G into consecutives bipartite digraphs of depth one. The scheduling obtained by the concatenation of the schedules of these bipartite digraphs is a feasible schedule or may be modified to a feasible schedule of G. Now, if we can determine the necessary and sufficient conditions to exist idle times in this feasible schedule then we can determine the complexity of this problem. This is a useful guide and foundation for future research.

I.